\documentclass{article}

\PassOptionsToPackage{numbers, compress}{natbib}

\usepackage[final]{neurips_2022_ml4ps}
\usepackage{amsmath}
\usepackage{graphicx}

\usepackage[utf8]{inputenc} 
\usepackage[T1]{fontenc}    
\usepackage{hyperref}       
\usepackage{url}            
\usepackage{booktabs}       
\usepackage{amsfonts}       
\usepackage{nicefrac}       
\usepackage{microtype}      
\usepackage{xcolor}         

\title{Amortized Bayesian Inference for Supernovae in the Era of the Vera Rubin Observatory Using Normalizing Flows}

\author{%
  V. Ashley Villar \\
  Department of Astronomy \& Astrophysics\\
  Pennsylvania State University\\
  State College, PA 16802 \\
  \texttt{vav5084@psu.edu} \\
  \\
  Institute for Computational \& Data Sciences\\
  Pennsylvania State University\\
  State College, PA 16802 \\
  \\
 Institute for Gravitation and the Cosmos\\
  Pennsylvania State University\\
  State College, PA 16802 \\
}

\begin{document}

\maketitle

\begin{abstract}
  The Vera Rubin Observatory, set to begin observations in mid-2024, will increase our discovery rate of supernovae to well over one million annually. There has been a significant push to develop new methodologies to identify, classify and ultimately understand the millions of supernovae discovered with the Rubin Observatory. Here, we present the first simulation-based inference method  using normalizing flows, trained to rapidly infer the parameters of toy supernovae model in multivariate, Rubin-like datastreams. We find that our method is well-calibrated compared to traditional inference methodologies (specifically MCMC), requiring only one-ten-thousandth of the CPU hours during test time. 
\end{abstract}

\section{Introduction}
Supernovae (SNe) mark the explosive deaths of stars, appearing in the night sky as rapidly-evolving thermal events in distance galaxies. Theoretically, the physics of these events can be fully described by observing the electromagnetic radiation as a function of both wavelength and time (assuming spherical symmetry). In practice, broadband photometry captures light integrated over $\sim100$ nm with a typical cadence of several days (see Figure 1). Standard Bayesian inference techniques (e.g., Markov chain Monte Carlo or nested sampling) are traditionally used to fit physical models to the observational data in order to derive the underlying parameters of the SNe and their progenitor stars. These methods, depending on the complexity of the model, can take between 10s of minutes to several days on a single CPU to sample the posterior.

In mid-2024 a new observatory, the Vera C. Rubin Observatory, will begin a ten year survey of the southern sky known as the Legacy Survey of Space and Time (LSST). LSST will increase our annual discovery rate of SNe from approximately 10k to well over 1 million. These will be buried within 10 million alerts reported \textit{nightly}. At these scales, classical inference techniques become prohibitively expensive. Here, we propose to utilize simulation-based inference (SBI) to rapidly estimate the posterior $p(\theta|y)$ of the SN physical properties, where $\theta$ represents the underlying model parameters of the SN and $y$ are the light curve observations. In particular, we explore utilizing amortized inference with masked autoregressive normalizing flows (MAFs) \cite{rezende2015variational, papamakarios2017masked}. Rather than evaluating the likelihood to estimate the posterior, the SBI method uses a large simulation set to train a normalizing flow with parameters $\phi$ to approximate $p(\theta|y)$.

SBI for rapid inference is growing increasingly popular in the fields of galaxy evolution \cite{hahn2022accelerated}, gravitational waves \cite{green2020gravitational} and cosmology \cite{hortua2020constraining}, among others. Despite this, there has been little work in developing SBI methods for SN science. Recently \cite{sanchezamortized} presented an amortized, stochastic variational inference method for SN light curves. However, this work was limited to to unphysical and simplified parametric models with univariate light curves (i.e., light curves observed in just one photometric filter). Here, we present the first attempt to perform SBI with normalizing flows on multivariate light curves and physical models.

\section{Data and Method}

\begin{figure}
  \centering
   \makebox[\textwidth][c]{\includegraphics[width=1.1\textwidth]{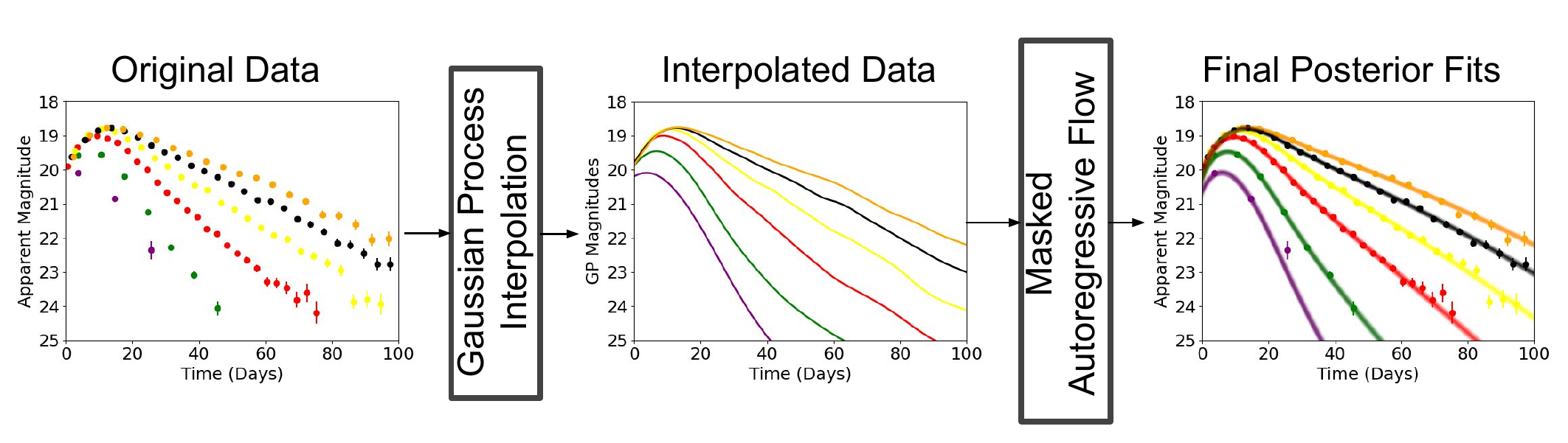}}%
  \caption{Schematic of proposed pipeline explored in this work. LSST-like simulated SN light curves are fed into a Gaussian process interpolator. The interpolated light curves are finally used to train a masked autoregressive flow to estimate the model parameter posteriors; 300 samples of the estimated posterior are shown for the example light curve and are in excellent agreement with the data. \label{fig:diagram}}
\end{figure}
In this section, we describe our simulations, pre-processing steps and inference pipeline; the pipeline is summarized pictorially in Fig.~\ref{fig:diagram}. 

We simulate a realistic dataset of ``stripped" core-collapse SNe powered by the radioactive decay of Ni$^{56}$ and Co$^{56}$. These events are described by the so-called Arnett model found by equating the radioactive material created in the SN to the work done by the SN ejecta and the SN luminosity (first presented in \cite{arnett1982type}). From this, one can numerically solve the spectral energy distribution of the explosion as a function in time. This has been utilized extensively in the literature (see e.g., \cite{chatzopoulos2012generalized, villar2017theoretical}). For our purposes, it is only important to understand that the flux of a SN, as a function of wavelength and time, is deterministic and a function of only the following properties: the distance to the SN (also known as its redshift), the mass of the SN ejecta, the velocity of the SN ejecta, the fraction of radioactive material and the time of explosion.

The simulated light curves are observed in 6 bands ($ugrizy$), ranging from near UV to the near IR. We use the ``rolling" cadence strategy proposed to LSST, described in \cite{lochner2018optimizing}; with this strategy, the multi-variate light curves are sampled at irregular time intervals, with $r-$, $i-$ and $z-$bands observed much more frequently than $u-$ and $g-$bands (see Fig.\ref{fig:diagram}). We inject SN light curves between 1-10 days prior to the first observation in order to ensure a rise time would be observed and uniformly between redshifts $0.01-0.1$. We note that these SNe would be most likely used for population-level studies, due to their high signal-to-noise ratios and the well-measured rise (often necessary for parameter estimation). We use the standard assumptions\footnote{https://smtn-002.lsst.io/} to calculate the signal-to-noise ratio of each observation, assuming Gaussian noise that is dependent entirely on the flux of the observation, the average sky brightness and properties of the Rubin Observatory. We assume that the redshift of each SN is well measured, in agreement with the expected photometric redshift standard provided by LSST (see e.g., \cite{graham2017photometric}. For the remaining four parameters, we assume the following flat priors:

\begin{equation}
  \begin{split}
    \log(f_\mathrm{Ni}) &\sim U(-1.5,-0.5)\\
    \log(M_\mathrm{ej}) &\sim U(-1,1)
  \end{split}
\quad  \quad
  \begin{split}
    \log(V_\mathrm{ej}) &\sim U(3.5,4.5)\\
    t_\mathrm{exp} &\sim U(-10,0)
  \end{split}
\end{equation}
, where $f_\mathrm{Ni}$ is the fraction of radioactive material ($^{56}$Ni) in the ejecta, $M_\mathrm{ej}$ is the ejecta mass, $v_\mathrm{ej}$ is the ejecta velocity and $t_\mathrm{exp}$ is the time of explosion.
We do not include effects of dust reddening. Each SN must be converted into fixed-length vectors for our MAF. For this, we interpolate each light curve by fitting a 2D Matern-3/2 Gaussian process (GP). The GP interpolates over both wavelength and time, allowing us to resample the light curves at a fixed time grid. This method has been used many times in the SN literature (see e.g., \cite{boone2019avocado,villar2020superraenn,villar2020anomaly,qu2021scone}). We simulate a total of 20,000 SNe to train our SBI pipeline; each SN is encoded into 601 features: 100 flux samples from the observer frame in each of the $ugrizy$ bands and the redshift of the object. We emphasize that the pre-processing steps are likely to impact the ability of the SBI method to \textit{exactly} reproduce the posterior obtained via traditional methods. The GP somewhat denoises the light curve by fitting it to a smooth function, and alternatively can occasionally overfit to the noisy observed data. We will explore this more in our Results.

With simulation in hand, we aim to train a SBI method to estimate the posteriors of the SN parameters given the observations (and redshift) and compare this SBI method to traditional Bayesian inference.  In our baseline comparison, we will use the affine invariant MCMC ensemble sampler implementation in the \texttt{Python} module \texttt{emcee} \cite{foreman2013emcee}, arguably the most common sampling technique used in the subfield. For a typical SN with $\sim90$ observational data points, it takes roughly $\sim10$ minutes on one CPU for 50 walkers to take 5,000 steps (at which point the MCMC is typically converged). 

Our SBI pipeline is implemented in the \texttt{Python} module \texttt{sbi}\cite{tejero-cantero2020sbi}. As stated, we use a MAF architecture to approximate the model posterior. In short, this network combines the classical inference technique of variational inference (which is traditionally limited to Gaussian posterior estimates) with normalizing flows. The flow allows for the network to learn a flexible posterior involving nonlinear transformations of the observables. In particular, the MAF architecture uses a series of stacked Masked Autoencoder for Distribution Estimation \cite{2015arXiv150203509G} (MADE) blocks to transform a simple Gaussian distribution (which the user can sample in test time) to an estimate of the true posterior, $p_\phi(\theta|y)$. The difference between the estimated ($p_\phi(\theta|y)$) and true posterior ($p(\theta|y)$) is quantified and minimized via the Kullback–Leibler divergence between the two distributions. We minimally tune hyperparameters of our network by hand, finding reasonable results for a model with 5 MADE blocks with 200 hidden units. We use a 9-1 training-to-validation split and optimize the flow using the standard \texttt{Adam} optimizer \cite{2014arXiv1412.6980K}.

\begin{figure}
  \centering
  \includegraphics[width=10cm]{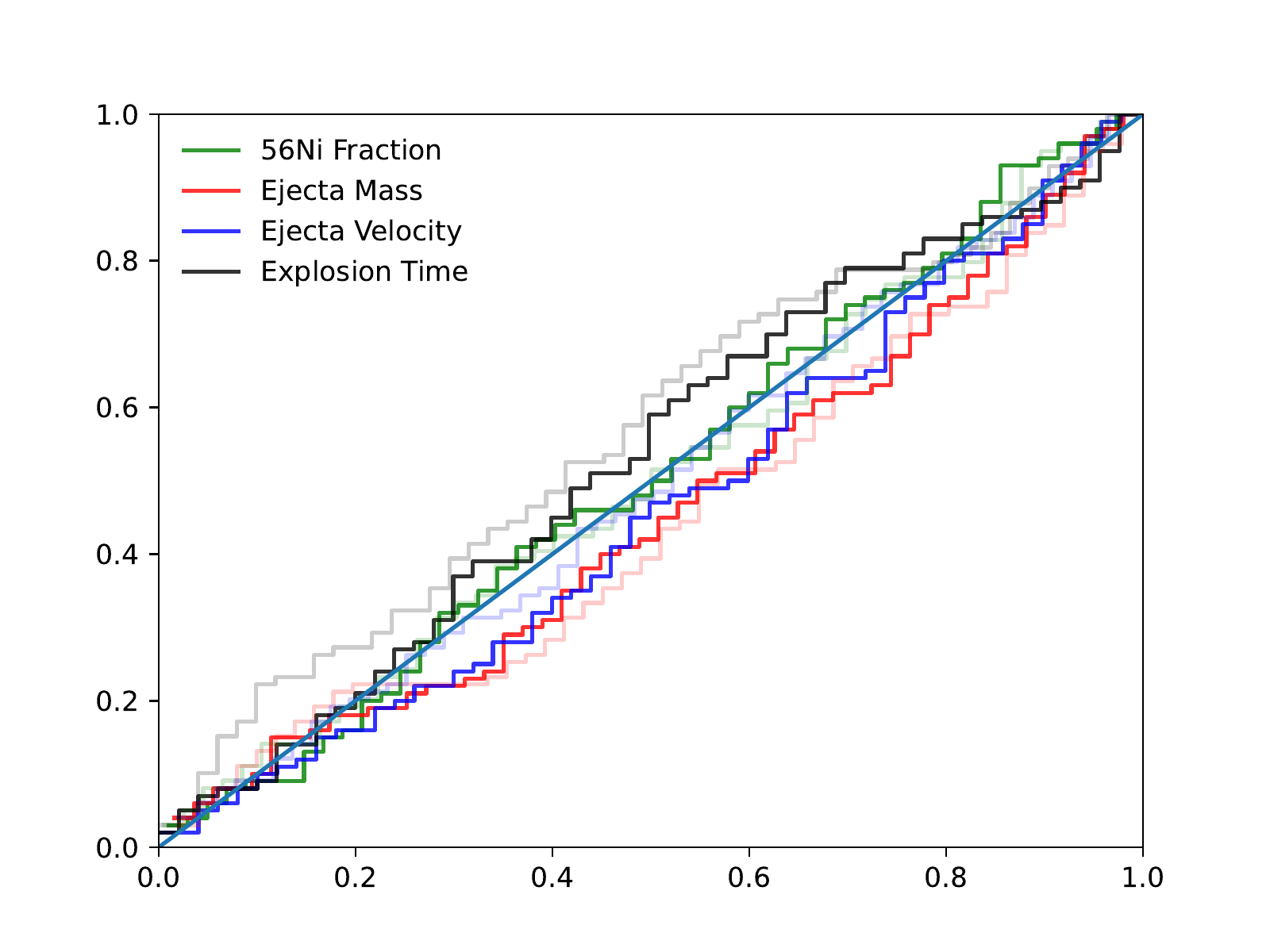}
  \caption{Probability-probability plot for the MCMC (light colors) and SBI methods (bold colors). For each of the four SN parameters, the colored curve represents the cumulative distribution of the percentile scores of the \textit{true} value of the parameter within the one-dimensional, marginalized posteriors. A perfectly calibrated posterior will follow the one-to-one diagonal (plotted in light blue) exactly. For both MCMC and SBI, 100 test light curves are used to generate the plot. Our SBI method is similar to the MCMC posteriors, indicating that our posteriors are well calibrated.\label{fig:pp}}
\end{figure}

\section{Results and Discussion}

\begin{figure}
  \centering
  \includegraphics[width=1\textwidth]{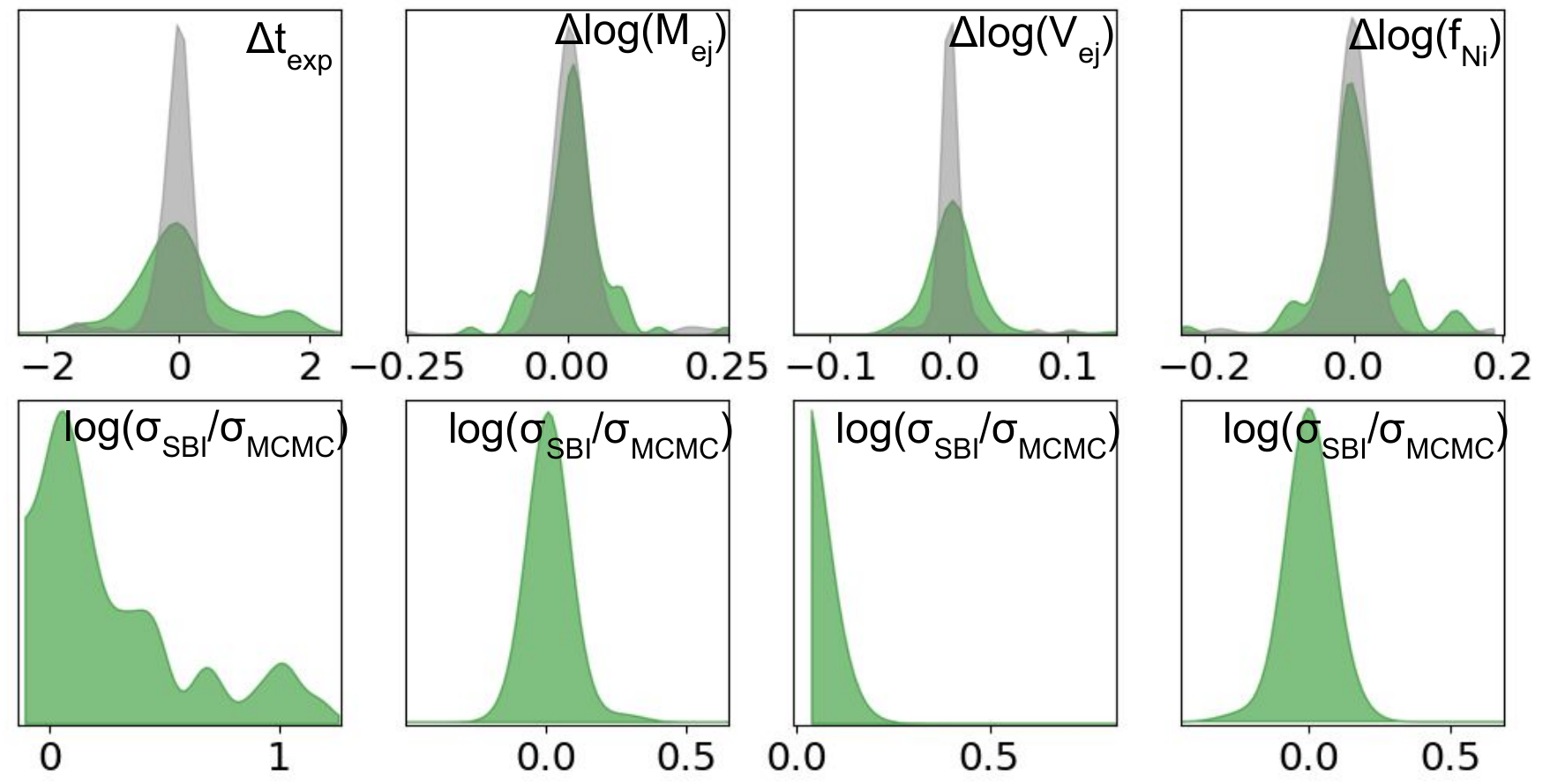}
  \caption{\textit{Top:}Kernel density estimate (KDE) of the distribution of the difference between the true model value and the median of the marginalized posterior for each of the four physical parameters, plotted for both MCMC (grey) and SBI (green) methods. \textit{Bottom:} KDE of the ratio of the standard deviation of each marginalized posterior as measured with SBI and MCMC. In general, the SBI method is unbiased compared to the MCMC method. SBI typically has somewhat larger uncertainties in explosion time and ejecta velocity, but is comparable for ejecta mass and fraction of radioactive material ($f_\mathrm{Ni}$). \label{fig:hist}}
\end{figure}
We use our trained pipeline to estimate the 4D posteriors of the SN parameters for the test set by simply drawing posterior samples from the conditional $p_\phi(\theta|x)$. For testing, we create a test set of 100 SN light curves, sampling from the same prior used to train the SBI pipeline. We fit these SNe with MCMC, and we estimate their corresponding posteriors with our SBI pipeline.

We first explore if the posteriors of the MAF are well calibrated -- i.e., if the percentiles' corresponding posterior cumulative distributions (CDFs) truly do reflect the probabilities of each set of model parameters given the data. In Figure~\ref{fig:pp} we show this ranking calibration. We specifically compare the posteriors derived by our baseline method, MCMC, and our SBI pipeline. The two methods result in similar calibrations, suggesting that our SBI posteriors are well-calibrated compared to traditional sampling methods.

We next explore how comparable our posteriors are to the baseline methodology. We note again that we do not necessarily expect the SBI posteriors to precisely match those of the baseline MCMC method -- even if they are well-calibrated. The SBI is trained on a heavily pre-processed version of the original data. In Figure~\ref{fig:hist}, we compare the difference between the medians of the marginalized posteriors as well as the ratio between the standard deviations of the marginalized posteriors, estimated with both MCMC and SBI. We find that our SBI posteriors are unbiased - largely matching the MCMC posteriors although with slightly more scatter. Corroborating this, we see that the SBI standard deviations are typically somewhat larger than those found with traditional MCMC. This is likely due to some additionally scatter apparently induced by the GP interpolation method. However, this seemingly most strongly impacts the time of explosion and ejecta velocity.

Finally, we compare the computational costs of both methods. As stated, for a typical light curve, the MCMC method takes roughly $\sim10$ minutes to converge. In contrast, in test time, the full SBI pipeline takes $\sim5-10$ms per object, with the majority of that time ($>5$ms) used to fit the GP interpolator. In training time, the SBI takes only $\simeq30$ minutes of CPU time to fully train.

In summary, we have presented the first SBI method for rapid parameter inference of multivariate SN light curves. We have demonstrated that our method is well-calibrated, resulting in believable posteriors. In future work, we plan to extend this algorithm to work with live streaming data in order to provide SN inference in real time, as well as to expand the various physical types of SNe modeled.

\section{Broader Impact}
The SBI pipeline presented here is an essential step in SN science given the millions of SNe expected to be discovered with LSST. In addition to enabling population-level statistical studies, SBI methods save valuable CPU time and conserve energy. We note that, as a word of caution, SBI methods like the one illustrated here can be well-calibrated for events within the simulated training set but will provide largely meaningless results for out-of-distribution events (e.g., a SN not well-described by the Arnett model in our case).

\bibliographystyle{IEEEtranN}
\bibliography{mybib.bib}


\end{document}